\documentstyle{mn}
\def\spose#1{\hbox to 0pt{#1\hss}}
\def\simlt{\mathrel{\spose{\lower 3pt\hbox{$\mathchar"218$}}
     \raise 2.0pt\hbox{$\mathchar"13C$}}}
\def\simgt{\mathrel{\spose{\lower 3pt\hbox{$\mathchar"218$}}
     \raise 2.0pt\hbox{$\mathchar"13E$}}}
\newcommand{\cntr}[1]{\multicolumn{1}{c}{#1}}

\begin{document}
\title[PDS~456: A radio-quiet analogue of 3C~273?]{Optical and
infrared observations of the luminous quasar PDS~456: A radio-quiet
analogue of 3C~273?}
\author[C. Simpson et al.]{Chris Simpson,$^1$ Martin Ward,$^2$, Paul
O'Brien,$^2$ and James Reeves$^2$\\
$^1$Subaru Telescope, 650 N. A`Oh\={o}k\={u} Place, Hilo, HI 96720,
U.S.A.\\
$^2$X-Ray Astronomy Group, Department of Physics and Astronomy,
University of Leicester, Leicester LE1 7RH}
\maketitle

\begin{abstract}\normalsize
We present infrared photometry and optical and infrared spectroscopy
of the recently-discovered, extremely luminous nearby quasar PDS~456.
A number of broad emission features are seen in the near-infrared
which we are unable to identify. We measure a more accurate redshift
from a narrow forbidden emission line and compare the
optical--infrared spectrum to that of 3C~273. The close similarity
suggests that PDS~456 is a radio-quiet analogue of 3C~273, although
radio observations do not support this idea.
\end{abstract}

\section{Introduction}

Torres et al.\ (1997; hereafter T97) reported the discovery of a new
bright ($V = 14.0$) quasar at relatively low redshift ($z =
0.184$). This object, called PDS~456, was discovered in the Pico dos
Dias survey for young stellar objects, which uses optical magnitude
and {\it IRAS\/} colours as selection criteria. Although somewhat
fainter than 3C~273, PDS~456 lies close to the Galactic centre and is
seen through $A_V \approx 1.5$\,mag of extinction (T97), and it is
therefore intrinsically more luminous. It is, however, radio-quiet,
with $S_{4.85} < 42$\,mJy (Griffith et al.\ 1994) and $S_{1.4} =
22.7$\,mJy (PDS~456 can be identified with NVSS~J172819$-$141555;
Condon et al.\ 1998), where $S_\nu$ is the flux density at $\nu$\,GHz.

In this paper we present optical and infrared observations of PDS~456.
We confirm the redshift found by T97, and make a more accurate
determination based on the unblended narrow emission line of
[Fe~II]~$\lambda$1.6435\,$\mu$m. We also confirm the existence of
significant Galactic reddening. We compare the properties of PDS~456
with those of 3C~273, with which it has a number of similarities, and
investigate the possibility that it is a radio-quiet analogue of the
latter source.

We adopt $H_0 = 50$\,km\,s$^{-1}$\,Mpc$^{-1}$ and $q_0 = 0.1$. With
this cosmology and our improved redshift determination, the proper
distance to PDS~456 is 1.01\,Gpc.

\section{Observations and reduction}

\subsection{Infrared imaging}

{\it JHKL$'$M\/} images of PDS~456 were obtained using IRCAM3 on the
United Kingdom Infrared Telescope (UKIRT) on UT 1997 August 24.
Standard near-infrared jittering techniques were used to allow
flatfielding without the need to observe blank regions of sky. The
dark-subtracted images were scaled to the same median pixel value and
median-filtered to produce a flatfield image which was divided into
the individual frames. These frames were then registered using the
centroid of the quasar and averaged with bad pixel masking. Total
integration times were 50\,s at {\it JHK\/}, 270\,s at $L'$, and 72\,s
at $M$. Conditions were photometric and flux calibration was performed
using a solution derived from observations of a number of standard
stars during the course of the night. Similar observations were also
performed with the same instrument on UT 1997 September 16.

\subsection{Infrared spectroscopy}

Near-infrared spectra of PDS~456 in the $J$ and $K$ windows were taken
using UKIRT/CGS4 on UT 1997 August 12 and 10, respectively. Conditions
were photometric throughout and both spectra have a resolution $R
\approx 900$ and total integration times of 640\,s. Bad pixel masking
and interleaving of the separate integrations was performed with the
Starlink CGS4DR software, and the remaining reduction was undertaken
with the IRAF data reduction package. The spectra were taken through a
1.2-arcsec slit and extracted along a single 1.2-arcsec pixel. The $J$
spectrum was corrected for atmospheric absorption and flux calibrated
using the A2 star HD~161903. The $K$ spectrum used the F0 star
SAO~121857 as an atmospheric dividing standard and was flux calibrated
using HD~225023. The spectra were wavelength calibrated using krypton
($J$-band) and argon ($K$-band) arc lamps and the r.m.s.\ deviation
from the adopted fit was $\simlt 1$\,\AA.

An 8--13\,$\mu$m spectrum of the source was taken with UKIRT/CGS3 on
UT 1997 August 30 as a service observation. A 5.5-arcsec aperture was
used with the low resolution grating, which gives a resolution $R
\approx 50$. Triple sampling was employed and the total integration
time was 3600\,s. A spectrum was taken of $\alpha$~Aql (Altair) at
similar airmass, which was used for flux calibration using its ratio
with the spectrum of $\alpha$~Lyr (Vega, assumed to be a 9400\,K
blackbody) from Cohen \& Davies (1995).

\subsection{Optical spectroscopy}

Optical spectra were taken with the Intermediate Dispersion
Spectrograph (IDS) on the 2.5-m Isaac Newton Telescope (INT) as
service observations on the night of UT 1997 August 22. Four grating
settings were used to cover the entire optical spectrum from
3320--9150\,\AA\ at a dispersion of $\sim 1.6$\,\AA\,pixel$^{-1}$. A
1.5-arcsec slit was used, and the seeing was reported to be
1.0\,arcsec. The total integration time at each position was 500\,s,
split into three separate exposures to facilitate cosmic ray
removal. Wavelength calibration was achieved through the use of argon
and neon lamps, and the r.m.s.\ deviation from the adopted fit was
$\simlt 0.1$\,\AA\ for all wavelength regions except in the range
$5800\,{\rm\AA} \simlt \lambda \simlt 6100$\,\AA, where the arc lines
were saturated, and the deviations became as large as 1.5\,\AA. We
determined the flux scale using Bohlin's (1996) spectrum of
BD~+33$^\circ$2642 (which was also observed with the same instrumental
setup), available from the Space Telescope Science Institute. The
three exposures at each grating position were averaged together after
masking out cosmic ray hits, and the spectra merged by using the
regions of overlap to compute necessary greyshifts between the
different grating positions. Shifts of up to 8\% were needed to match
up the flux scales, and it was noted that the fluxes in the different
exposures taken at a single grating position differed by as much as
25\%. This strongly suggests the presence of clouds or significant
slit losses, and as such the flux scale must be considered
approximate.

\section{Results and analysis}

\subsection{Photometry}

The results of our aperture photometry are presented in
Table~\ref{tab:photom}. The flux calibration solutions for the night
of 1997 Sep 16 were slightly more uncertain than for the night of 1997
Aug 24, but we find no evidence for variability in any of the five
filters at greater than $2\sigma$ significance. This supports the lack
of variability observed by T97 at optical wavelengths over a period of
three weeks.

\begin{table}
\caption[]{Infrared photometry of PDS~456 in a 5-arcsec aperture.}
\label{tab:photom}
\centering
\begin{tabular}{cccc}
Filter & Central & \multicolumn{2}{c}{Flux (mJy)} \\
& wavelength ($\mu$m) & 1997 Aug 24 & 1997 Sep 16 \\ \hline
$J$  & 1.25 &  $24.9 \pm  0.6$  &  $27.0 \pm  1.8$ \\
$H$  & 1.65 &  $37.0 \pm  1.6$  &  $40.9 \pm  2.7$ \\
$K$  & 2.20 &  $72.3 \pm  2.5$  &  $82.8 \pm  5.2$ \\
$L'$ & 3.80 & $184.2 \pm 15.5$  & $176.0 \pm 16.6$ \\
$M$  & 4.80 & $251.3 \pm 20.2$  & $199.6 \pm 30.6$ \\
\hline
\end{tabular}
\end{table}

A power-law fit to the observed $H - K$ colour (and $K$-band continuum
from our spectrum) gives $\alpha = 2.4 \pm 0.1$ ($S_\nu \propto
\nu^{-\alpha}$). This is very steep for quasars, which have typical
spectral indices $\alpha = 1.4 \pm 0.3$ (Neugebauer et al.\ 1987), and
suggests significant reddening is present. The near-infrared colours
imply $A_V = 1.5 \pm 1.0$, in line with the estimate of T97.

\subsection{Spectroscopy}

\begin{figure}
\includegraphics{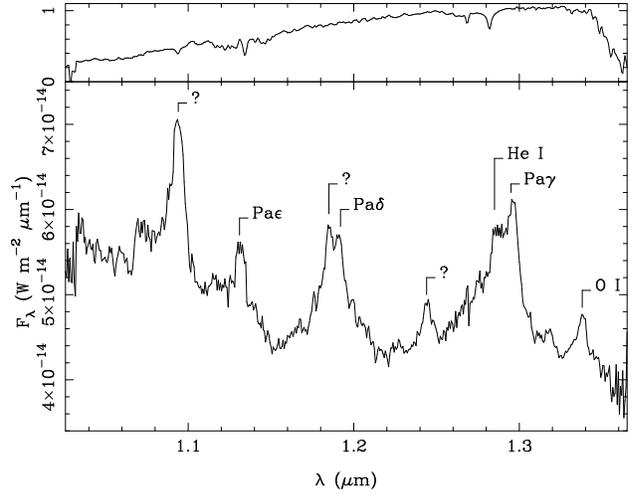}
\vspace*{63mm}
\caption[]{Observed $J$-band spectrum of PDS~456. The top trace shows
the normalized spectrum of the standard star HD~161903, revealing the
locations of atmospheric and stellar absorption. Although the feature
at 1.095\,$\mu$m may be contaminated by poor removal of Pa$\gamma$
absorption from the dividing standard, its equivalent width is clearly
too large, and there may be emission in addition to Pa$\zeta$.}
\label{fig:jspek}
\end{figure}
 
\begin{figure}
\includegraphics{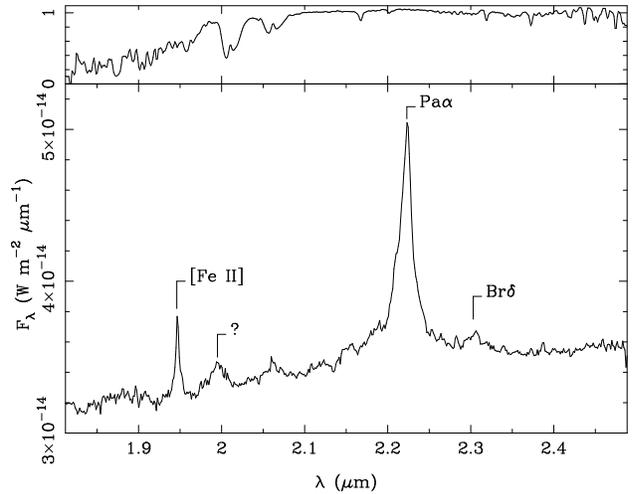}
\vspace*{63mm}
\caption[]{Observed $K$-band spectrum of PDS~456. The top trace shows
the normalized spectrum of SAO~121857. Note how the weak broad feature
near 2\,$\mu$m is at the wrong wavelength to be imperfectly removed
CO$_2$ absorption.}
\label{fig:kspek}
\end{figure}

\begin{figure*}
\includegraphics{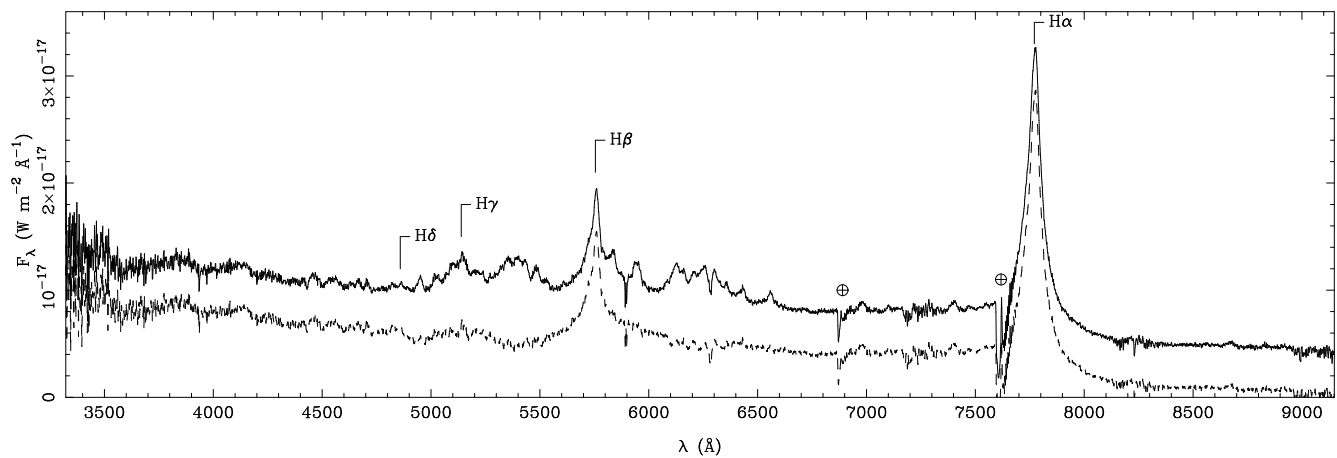}
\vspace*{70mm}
\caption[]{Optical spectrum of PDS~456. The Balmer lines are marked,
and atmospheric absorption features are indicated by a $\oplus$
symbol. The dashed line is the same spectrum after removal of the
optical Fe~II emission (see text for details), and shifted downwards
for clarity.}
\label{fig:optspek}
\end{figure*}

We present our near-infrared spectra in Figs~\ref{fig:jspek} and
\ref{fig:kspek}. We confirm the redshift found by T97 on the basis of
the [Fe~II]~$\lambda$1.6435\,$\mu$m emission line, the only unblended
forbidden line in the entire spectrum. The sharpness of this line
(${\rm FWHM} \approx 1000$\,km\,s$^{-1}$) allows a more accurate
redshift determination than from the broad lines, and we measure $z =
0.18375 \pm 0.00030$. The optical spectrum is shown in
Fig.~\ref{fig:optspek}.

There are a number of emission features in our near-infrared spectra
which we are unable to identify. Their rest wavelengths (assuming they
are at the redshift of PDS~456) do not correspond to lines seen in the
spectra of other quasars or Seyfert galaxies (e.g.\ Hill, Goodrich \&
DePoy 1996; Thompson 1995). However, they also do not correspond to
features in the dividing standards, whose spectra we also show in
Figs~\ref{fig:jspek} and \ref{fig:kspek}. The emission features
observed at 1.185\,$\mu$m and 1.245\,$\mu$m are the most clearly real,
since the emission feature observed at 1.095\,$\mu$m may be composed
partly of Pa$\gamma$ from the dividing standard and Pa$\zeta$ from the
quasar. However, the former line is in a fairly clean part of the
atmosphere and was readily removed, and Pa$\zeta$ should be weaker
than Pa$\epsilon$. It therefore appears that there may be another
unidentified emission line at this wavelength. The weak broad feature
near 2\,$\mu$m is at the wrong wavelength to be caused by
imperfectly-subtracted atmospheric CO$_2$ absorption, and again we
suspect it is a real emission feature in the quasar spectrum. We have
re-reduced the $J$-band spectrum with a dividing standard of
completely different spectral type (K0III), and we also re-observed
PDS~456 in the $J$ window with a different camera/grating combination
of CGS4 on UT 1998 August 26. Both times the final spectrum was
indistinguishable from Fig.~\ref{fig:jspek}. The wavelengths of the
lines are inconsistent with their being lines from higher orders
(e.g.\ H$\alpha$), and their relative wavelengths do not correspond to
any pair of strong lines, so we can rule out their being from a single
system at a different redshift (either lower or higher).  We obviously
cannot conclusively rule out their being more than one additional
system along the line of sight, although this is very unlikely,
especially given our small spectroscopic aperture.

The H~I lines have such broad wings (${\rm FWZI} \simgt
30\,000$\,km\,s$^{-1}$) that fluxes are difficult to measure
reliably. In addition, many of the lines are blended, further
hampering measurements of their fluxes. We have opted to use the
Pa$\alpha$ line as a template for measuring the fluxes of the blended
H~I lines. We first subtract a low-order cubic spline from our
$K$-band spectrum, masking out regions contaminated by emission lines,
and interpolating across the Br$\delta$ line. We then progressively
subtract a scaled version of the Pa$\alpha$ line from the locations of
the other emission lines until the resultant spectrum shows no
evidence of line emission. We determine the strength of the other line
in the blend by measuring the flux above an adopted continuum level in
the spectrum with the hydrogen line subtracted. In the case of
He~I~$\lambda$1.0830\,$\mu$m (blended with Pa$\gamma$), this results
in a very broad line (${\rm FWHM} \approx 7000$\,km\,s$^{-1}$,
compared to ${\rm FWHM} \approx 3500$\,km\,s$^{-1}$ for the H~I lines)
with a pronounced blue wing, as can be anticipated from
Fig.~\ref{fig:kspek}. While no strong emission lines have been
observed in this wavelength region in other objects, we cannot rule
out the possibility that the He~I line is further blended because of
the presence of the unidentified emission lines in our spectrum. We
note, however, that Netzer (1976) suggests that the He~I lines should
be broader than those of H~I.

A determination of the H$\beta$ flux is further complicated by the
strong Fe~II emission which produces a ``pseudo-continuum'' on both
sides of the emission line. We removed the Fe~II emission from the
spectrum by first performing a fit (by eye) to the emission lines so
that the continuum was approximately linear over short wavelength
intervals. A Gaussian fit to the isolated 4177\,\AA\ and 5534\,\AA\
lines was used to model the profiles of the lines. We then took each
individual line in turn and determined the flux which produced the
minimum sum-of-squares residual about a continuum level which was
linearly interpolated between points either side of the line. A new
spectrum was constructed using these line fluxes and the process
repeated until the result converged. The Fe~II-subtracted spectrum is
shown in Fig.~\ref{fig:optspek}.

\begin{table}
\caption[]{Observed line fluxes and rest-frame equivalent widths. The
major uncertainty in the line fluxes comes from continuum placement.
The fluxes from the $K$-band spectrum have been increased
by 25\% (see text for details).}
\label{tab:fluxes}
\centering
\begin{tabular}{llr@{}l@{$\,\pm\,$}r@{}lr}
Line & \cntr{$\lambda_{\rm rest}$} & \multicolumn{4}{c}{Flux}
& \cntr{Equivalent} \\
& \cntr{($\mu$m)} & \multicolumn{4}{c}{($10^{-17}$\,W\,m$^{-2}$)}
& \cntr{width (\AA)}
\\ 
\hline
H$\gamma$    & 0.4340 &  28 &    &  8 &    &  28.2 \\
H$\beta$     & 0.4861 & 110 &    & 15 &    &  93.2 \\
H$\alpha$    & 0.6563 & 360 &    & 30 &    & 503.1 \\
Pa$\zeta$+?  & 0.924  &  20 &    &  3 &    &  33.3 \\
Pa$\epsilon$ & 0.9552 &   6 &    &  1 &    &  10.8 \\
?            & 0.999  &  23 &    &  5 &    &  44.9 \\
Pa$\delta$   & 1.0052 &  14 &    &  3 &    &  28.1 \\
?            & 1.051  &   3 & .6 &  0 & .6 &   6.8 \\
He~I         & 1.0830 &  26 &    &  6 &    &  50.8 \\
Pa$\gamma$   & 1.0941 &  24 &    &  5 &    &  46.8 \\
O I          & 1.129  &   5 & .1 &  0 & .5 &  10.1 \\
{}[Fe~II]    & 1.6435 &   4 & .5 &  0 & .5 &   9.3 \\
?            & 1.684  &   4 & .2 &  0 & .6 &   8.4 \\
Pa$\alpha$   & 1.8756 &  62 &    &  9 &    & 130.1 \\
Br$\delta$   & 1.9451 &   2 & .7 &  0 & .8 &   7.3 \\
\hline
\end{tabular}
\end{table}

From this analysis, we measure the flux of the Fe~II 37,38 multiplets
as $F(\lambda4570) = (5.1 \pm 0.7) \times 10^{-16}$\,W\,m$^{-2}$ and
of the H$\beta$ line as $F({\rm H}\beta) = (1.1 \pm 0.2) \times
10^{-15}$\,W\,m$^{-2}$, three times larger than that determined by
T97. We believe this discrepancy is due to an incorrect placement of
the continuum level by T97 when performing their simple flux
determination. The value we obtain from our more detailed method is
likely to be far more accurate. The line fluxes are listed in
Table~\ref{tab:fluxes}. Note that for this Table and future
discussions, we have brightened our $K$-band spectrum by 25\% to match
our photometry, suspecting poor seeing and slit losses for the
difference; no such shift was needed for the $J$-band spectrum. The
equivalent width of H$\beta$ is fairly typical of nearby luminous
quasars (Miller et al.\ 1992) and the ratio
Fe~II~$\lambda$4570/H$\beta \approx 0.5$ is not unusually strong.

We also have reason to suspect T97's [O~III]~$\lambda$5007 flux, which
is blended with the $\lambda$5018 line of Fe~II multiplet 42, and
which we fail to detect in our spectrum after removing the Fe~II and
H$\beta$ emission as described above. T97 deblend these two lines and
obtain Gaussians of similar, narrow (${\rm FWHM} \approx
700$\,km\,s$^{-1}$), widths whose wavelengths are significantly
different from those expected. The mean wavelength of T97's two
deblended components is very close to the wavelength expected for the
$\lambda$5018 line alone (our revised redshift improves the
agreement), and their combined flux is in excellent agreement with the
flux we determine for this line alone. The Fe~II lines are also known
to be broad (we measure ${\rm FWHM} \approx 1500$\,km\,s$^{-1}$),
casting doubt on the narrow line produced by T97's deblending. We also
fail to detect [O~III]~$\lambda$4959 at its expected flux level. We
place an upper limit of $F({\rm[O~III]}~\lambda5007) < 2 \times
10^{-17}$\,W\,m$^{-2}$, corresponding to a rest-frame equivalent width
of $< 2$\,\AA.

\subsection{Reddening}

\begin{figure}
\vspace*{54mm}
\includegraphics{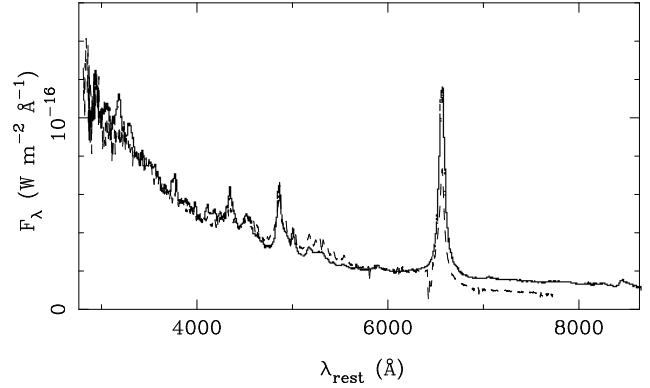}
\caption[]{Comparison of the rest-frame optical spectra of PDS~456
(dereddened by $A_V = 1.4$\,mag, dashed line) and 3C~273 ( solid
line). The spectrum of 3C~273 has been taken from Morris \& Ward
(1988) and the flux has been multiplied by a factor 1.77 to match the
flux of PDS~456 at $\lambda \approx 6000$\,\AA.}
\label{fig:3c273}
\end{figure}

Whilst the H~I line ratios can be used to estimate the reddening, they
are too uncertain to provide an accurate value, although they are
broadly consistent with the $A_V = 1.5$\,mag determined by T97 from
the equivalent width of the Na~D~1 line, and the extinction maps of
Burstein \& Heiles (1982). We make an additional estimate of the
extinction by comparing the optical spectrum of PDS~456 to that of the
quasar 3C~273. As Fig.~\ref{fig:3c273} shows, there is excellent
agreement (at least blueward of H$\alpha$; we discuss the disagreement
at longer wavelengths in the next section) when PDS~456 is dereddened
by $A_V = 1.4$. Since 3C~273 is itself reddened by $A_V \approx
0.1$\,mag (Burstein \& Heiles 1982), we infer a total Galactic
extinction of 1.5\,mag towards PDS~456, and adopt this value in
our later analysis.

\section{Discussion}

\begin{figure}
\includegraphics{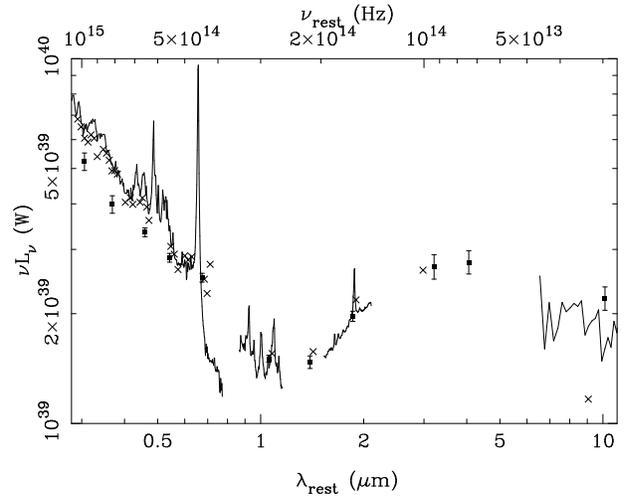}
\vspace*{63mm}
\caption[]{Dereddened (by $A_V=1.5$\,mag) optical-infrared spectra and
photometry (solid squares with error bars) of PDS~456, plotted in the
rest frame. The spectra have been smoothed and rebinned to improve
presentation. The {\it UBVRI\/} photometry is taken from T97. Also
plotted is the 12\,$\mu$m {\it IRAS\/} measurement, although this is
likely to include a contribution from the host galaxy and so the
discrepancy between it and our CGS3 spectrum is probably real. The
crosses are measurements of 3C~273 (dereddened by $A_V = 0.12$\,mag
and scaled by a factor 1.77 as in Fig.~\ref{fig:3c273}) taken from
Neugebauer et al.\ (1987; optical data) and Robson et al.\ (1986;
infrared data).}
\label{fig:optir}
\end{figure}

In the previous section, we noted the similarity between the optical
spectra of PDS~456 and 3C~273. In Fig.~\ref{fig:optir}, we present the
dereddened optical--infrared spectral energy distributions of the two
quasars. We use the data of Rieke \& Lebofsky (1985) to correct the
data at $\lambda > 3.5\,\mu$m, which is beyond the range of Cardelli
et al.'s (1993) extinction law. The similarity is once again striking,
with the exception of the 10\,$\mu$m flux, which is nearly a factor of
two higher in PDS~456 than 3C~273, relative to the
optical--near-infrared data. Note that although these data are
measured through different apertures, the quasar is more than 300
times brighter than an $L^*$ galaxy in the $K$-band (Gardner et al.\
1997), and so the stellar contribution will be negligible.

There is clearly a problem with the continuum level redward of
H$\alpha$, since it does not match the interpolation between the rest
of the optical spectrum and the $J$-band spectrum. This is not a
simple flux calibration error affecting the reddest of the four INT
sub-spectra which can be corrected with a constant shift, since the
match is excellent in the region of overlap with the adjoining
subspectrum.

The dereddened optical continuum is well-described by a power law with
spectral index $\alpha = -0.11 \pm 0.11$. This is bluer than the mean
optical spectral index for quasars, but still within the observed
range (Neugebauer et al.\ 1987). The inflexion at $\lambda_{\rm rest}
\sim 1.2\,\mu$m is a ubiquitous feature of quasar SEDs (Neugebauer et
al.\ 1987; Elvis et al.\ 1994). However, the near-infrared bump
usually extends to longer wavelengths, a power law often being able to
fit the SED throughout the range 1--10\,$\mu$m (see Neugebauer et al.\
1987).

Given the similarities between the optical and infrared properties of
PDS~456 and 3C~273, it is natural to ask the question: Is PDS~456 a
radio-quiet analogue of 3C~273? Certainly the low equivalent widths of
the optical forbidden lines and blue optical continuum suggest
blazar-like properties. A blazar nature should also be apparent at
radio wavelengths, since the core emission should be strongly boosted
by Doppler beaming (Falcke, Sherwood \& Patnaik 1996), resulting in a
flat radio spectrum which is more luminous than the general
radio-quiet quasar population. To investigate this possibility, we
have obtained VLA A-array data at 1.4 and 4.85\,GHz. A detailed
discussion of the data will be presented in a future paper, but they
confirm the NVSS identification and flux, and reveal a steep radio
spectrum. By extrapolating this spectrum to 8.4\,GHz, we predict a
flux of 4.6\,mJy, which would cause PDS~456 to lie on the
optical--radio luminosity relation for RQQs of Kukula et al.\ (1998).

The high optical luminosity and low forbidden-line equivalent widths
of PDS~456 are characteristic of the `Baldwin effect' (Baldwin 1977).
Although several explanations for this effect have been advanced,
including Doppler-boosting (Browne \& Murphy 1987), radiation from an
accretion disc (Netzer 1987), and reddening (Baker 1997), these all
incorporate an orientation dependence such that very luminous, very
low equivalent width sources like PDS~456 should be seen nearly
pole-on. The lack of a dominant, boosted radio core is therefore
something of a mystery.

\section{Summary}

We have presented optical and infrared spectra of the nearby luminous
quasar PDS~456. We measure a redshift $z = 0.18375 \pm 0.00030$ based
on the forbidden line of [Fe~II]~$\lambda$1.6435\,$\mu$m, but do not
detect any other forbidden emission lines. We detect at least three
emission lines in the near-infrared which we are unable to identify.
The dereddened optical continuum is rather blue and very similar to
that of 3C~273. Despite the similarities at optical wavelengths,
observations reveal that PDS~456 does not possess a strongly
Doppler-boosted radio core. We defer detailed discussion of the radio
observations to a later paper, wherein we will also present and
discuss an X-ray spectrum of PDS~456.

\section*{Acknowledgments}

The United Kingdom Infrared Telescope is operated by the Joint
Astronomy Centre on behalf of the U. K. Particle Physics and Astronomy
Research Council, and was ably piloted by Thor Wold. The Isaac Newton
Telescope is operated on the island of La Palma by the Isaac Newton
Group in the Spanish Observatorio del Rogue de los Muchachos of the
Instituto de Astrofisica de Canarias. The Very Large Array is part of
the National Radio Astronomy Observatory, which is operated by
Associated Universities, Inc., under cooperative agreement with the
National Science Foundation. The authors thank the staff at UKIRT and
the INT for taking the service observations presented in this paper,
and Duncan Law-Green for reducing the VLA data.


\begin{thebibliography}{}

\bibitem{}Baker, J. C., 1997, MNRAS, 286, 23

\bibitem{}Baldwin J. A., 1977, ApJ, 214, 679

\bibitem{}Bohlin R. C., 1996, AJ, 111, 1743

\bibitem{}Browne I. W. A., Murphy D. W., 1987, MNRAS, 226, 601

\bibitem{}Burstein D., Heiles C., 1982, AJ, 87, 1165

\bibitem{}Cardelli J. A., Clayton G. C., Mathis J. S., 1989, ApJ, 345,
245

\bibitem{}Cohen M., Davies J. K., 1995, MNRAS, 276, 712

\bibitem{}Condon J. J., Cotton W. D., Greisen E. W., Yin Q. F., Perley
R. A., Taylor G. B., Broderick J. J., 1998, AJ, 115, 1693

\bibitem{}Elvis M., et al., 1994, ApJS, 95, 1

\bibitem{}Falcke H., Sherwood W., Patnaik A. R., 1996, ApJ, 471, 106

\bibitem{}Gardner, J. P., Sharples, R. M., Frenk, C. S., Carrasco,
B. E. 1997, ApJ, 490, L99

\bibitem{}Griffith M. R., Wright A. E., Burke B. F., Ekers R. D., 1994,
ApJS, 90, 179

\bibitem{}Hill G. J., Goodrich R. W., DePoy D. L., 1996, ApJ, 462, 163

\bibitem{}Kukula M. J., Dunlop J. S., Hughes D. H., Rawlings S., 1998,
MNRAS, 297, 366

\bibitem{}Miller P., Rawlings S., Saunders R., Eales S., 1992, MNRAS, 254,
93

\bibitem{}Morris S. L., Ward M. J., 1988, MNRAS, 230, 639

\bibitem{}Netzer H., 1976, ApJ, 219, 822

\bibitem{}Netzer H., 1987, MNRAS, 225, 55

\bibitem{}Neugebauer G., Green R. F., Matthews K., Schmidt M., Soifer
B. T., Bennett J., 1987, ApJS, 63, 615

\bibitem{}Rieke G. H., Lebofsky M. J., 1985, ApJ, 288, 618

\bibitem{}Robson E. I., Gear W. K., Brown L. M. J., Courvoisier
T. J.-L., Smith M. G., Griffin M. J., Blecha A., 1986, Nature, 323,
134

\bibitem{}Thompson R. I., 1995, ApJ, 445, 700

\bibitem{}Torres C. A. O., Quast G. R., Coziol R., Jablonski F., de la
Reza R., L\'{e}pine J. R. D., 1997, ApJ, 488, L19 (T97)

\end{thebibliography}
\end{document}